\documentclass[12pt]{article}
\pdfoutput=1
\usepackage[affil-it]{authblk}
\usepackage[english]{babel}
\usepackage{amsmath, amsthm, amssymb}
\usepackage{enumerate}
\usepackage[left=2cm,right=2cm,top=2cm,bottom=2cm,a4paper]{geometry}
\usepackage{graphicx}
\usepackage{mathrsfs}
\usepackage{braket}
\usepackage{upgreek}
\usepackage{tensor}
\usepackage{slashed,cancel}
\usepackage{caption}
\usepackage{enumitem}
\usepackage{floatrow}
\usepackage{simplewick} 
\usepackage{comment}
\usepackage[onehalfspacing]{setspace}

\usepackage{tikz}
\DeclareGraphicsExtensions{.pdf,.png,.jpg}
\usepackage{fancyhdr}
\usepackage{cite}
\usepackage{bbm}                
\usepackage{xspace}				
\usepackage{hyperref}			
\hypersetup{
pdfstartview=FitV,
colorlinks=true, 
linkcolor=blue, 
citecolor=red, 
filecolor=black, 
urlcolor=blue}

\usepackage{subfigure}

\newcommand{\fb}{$\textrm{fb}^{-1}\ $}
\newcommand{\R}{$\mathcal{R}$}
\newcommand{\BR}{\mathrm{Br}}

\newcommand{\MET}{$\cancel{E}_{T}$}

\allowdisplaybreaks

\begin{document}

\setcounter{footnote}{0}
\setcounter{figure}{0}
\setcounter{table}{0}


\title{\textbf{Explaining the Lepton Non-universality at the  LHCb and CMS within a Unified Framework}}

\author[1,2]{Sanjoy Biswas\thanks{sanjoy.biswas@roma1.infn.it}}
\author[2]{Debtosh Chowdhury\thanks{debtosh.chowdhury@roma1.infn.it}}
\author[3,4]{Sangeun Han\thanks{gaeaearth@kaist.ac.kr}}
\author[3,5]{\\ Seung J. Lee\thanks{sjjlee@kaist.ac.kr}}

\affil[1]{\small Dipartimento di Fisica, Universit\`a di Roma La Sapienza, \authorcr {\it Piazzale Aldo Moro 2, I-00185 Roma, Italy}}
\affil[2]{\small Istituto Nazionale di Fisica Nucleare, Sezione di Roma, \authorcr {\it Piazzale Aldo Moro 2, I-00185 Rome, Italy}}
\affil[3]{\small Department of Physics, Korea Advanced Institute of Science and Technology, \authorcr {\it 335 Gwahak-ro, Yuseong-gu, Daejeon 305-701, Korea}}
\affil[4]{\small Center for Theoretical Physics of the Universe, IBS, Daejeon, Korea}
\affil[5]{\small School of Physics, Korea Institute for Advanced Study, Seoul 130-722, Korea}

%
%
%
%




\maketitle

\begin{abstract}
\noindent
The recent results from the LHCb in the context of $(B^+ \rightarrow K^+ l l)$ decay and the CMS analysis in the context of right handed $W$-boson ($W_R$) search show a $2.6\sigma$ and a $2.8\sigma$ deviations from the Standard Model expectations respectively. In this work, we address these two seemingly uncorrelated results in the context of \R-parity violating supersymmetry. We found that a particular combination of $LQD^c$-type operators which successfully explain the LHCb result, can also accommodate the CMS excess in the $eejj$ channel of the $W_R$ search.

\end{abstract}

\clearpage

\section{Introduction}\label{sec:intro}
The recent LHCb measurement has observed \cite{Aaij:2014ora} a significant deviation from the standard model (SM)
expectation of the ratio $R_K$, defined as $R_K = \BR(B^+ \rightarrow K^+ \mu^+ \mu^-)/\BR(B^+ \rightarrow K^+ e^+ e^-)$ \cite{Hiller:2003js}. 
The measurement predicts a $2.6\sigma$ deviation from the SM prediction, in the low invariat mass region ($1\ \textrm{GeV}^2 \leq M_{\ell\ell} \leq 6\ \textrm{GeV}^2$) of the di-lepton pair using a data set of 3 \fb integrated luminosity.

More interestingly, CMS analysis for the right-handed $W$-boson ($W_R$) search has also come up with a significant
deviation from the standard model expectation\cite{Khachatryan:2014dka}. The 
CMS search uses $pp$ collision data at the Large Hadron Collider (LHC)
at a center of mass energy of $8$ TeV with 19.7 \fb integrated luminosity. The invariant mass 
distribution $M_{eejj}$ shows an excess around 2 TeV, with a local significance of $2.8\sigma$ 
\cite{Khachatryan:2014dka}. The CMS collaboration has also reported a possible excess in the context of the di-leptoquark
search \cite{CMS:2014qpa}. The optimization of the data assuming a leptoquark mass $\sim$ 650 GeV yields a local significance of $2.4\sigma$ ($2.6\sigma$) in the $eejj$ ($e\nu jj $) channel.

There have already been quite a few studies in an attempt to explain these results separately assuming 
different models. In Ref. \cite{Hiller:2014yaa}, the authors studied the observed value of $R_K$ in the context of effective operator approach, illustrated with two leptoquark models. They also mentioned that the leptoquark couplings considered there, can correspond to certain \R-parity violating (RPV)  supersymmetric scenario. However the flavor structure of those  couplings can not address the CMS $eejj$ excesses. Certain constraints have been put on these effective operators in  \cite{Ghosh:2014awa}. On the other hand, the CMS excess in the context of both $W_R$ and di-leptoquark search is interpreted in \cite{Bai:2014xba} with a resonant coloron production and further decay of the coloron into a pair of leptoquark and 
in \cite{Dobrescu:2014esa}, with resonant production of vector like leptons via $W^{'}/Z^{'}$ vector boson.  A similar analysis \cite{Aguilar-Saavedra:2014ola} with $W^{'}/Z^{'}$ has been performed in the context of $W_R$ search. Refs. \cite{Deppisch:2014qpa,Heikinheimo:2014tba} showed that the $W_R$ excess can be explained within the context of GUT models. Within the framework of $\mathcal{R}$-parity violating supersymmetry (SUSY) an explanation via resonant slepton production has been provided in \cite{Allanach:2014lca} in the context of CMS $W_R$ search. In ref. \cite{Chun:2014jha} the di-leptoquark excess is explained.

 
Though it is quite preliminary to jump into any conclusion before a more detailed analysis of the data and despite the fact that the 
statistics is very low in the case of CMS analyses, one can still take these deviations at their face value in order to ensure a better search 
strategies either to claim a discovery or to put an exclusion limit. 
It is worth noting that while there are individual explanations for each of these results mentioned above, so far there has been no attempt to have explain both simultaneously.
In this article we present a unified framework which can accommodate both the LHCb and the CMS $W_R$ search results in the context 
of the RPV minimal supersymmetric standard model (MSSM). Given the fact that the deviation in the measured value of $R_K$ can be consistent with new physics (NP) either 
in the electron or in the muon sector due to the large theoretical uncertainties present in the SM expectations, in this article, we focus on the former possibility motivated by the observed CMS excess. 

The remaining of the paper is organized as follows: In section \ref{sec:model}, we give a brief account of the RPV model. Section \ref{sec:phenos} describes both the $B$-physics and collider consequences of this model. Finally, we summarize our results and conclude in section \ref{sec:conclusions}.

\section{Model: RPV SUSY}\label{sec:model}
In this section, we will give a brief review of $\mathcal{R}$-parity violating SUSY scenario.
While \R-parity conserving SUSY has many judicious features, which made it one of the most popular frameworks, \R-parity violating SUSY \cite{Hall:1983id,Ross:1984yg,Barger:1989rk,Dreiner:1997uz,Bhattacharyya:1997vv} provides an alternative. It can relax the naturalness bound from LHC due to the absence of large missing transverse energy (\MET) signature and at the same time provides rich collider phenomenology. In MSSM, the RPV interactions are generated through the following superpotential,
\begin{align}\label{eq:WRPV}
\mathcal{W}_{\rm hTL} =& \frac{1}{2}\lambda_{ijk}  L_i L_j E^c_k +\lambda^\prime_{ijk}  L_i Q_j D^c_k + 
\frac{1}{2}\lambda^{\prime\prime}_{ijk} U^c_i D^c_j D^c_k \,, \nonumber \\
\mathcal{W}_{\rm hBL} =& \mu^{\prime}_i L_i H_u \,,
\end{align}
where $L_{i}$, $E^{c}_{i}$ denote SU$(2)_L$ doublet and singlet superfields for leptons respectively, $Q_{i}$, $U^{c}_{i}$ and $D^{c}_{i}$ represent the left-handed quark doublet, right-handed up-type quark singlet and right-handed down-type quark singlet superfield respectively and $H_{u}$ is the up-type Higgs superfield that gives mass to the up-type quarks. Here, $\mathcal{W}_{\rm hTL}$ are the trilinear terms which contains only dimensionless parameters, and $\mathcal{W}_{\rm hBL}$ denotes the holomorphic bilinear terms containing dimensionful couplings. The $\lambda_{ijk}$'s and $\lambda^{'}_{ijk}$'s are Lepton number violating and $\lambda^{''}_{ijk}$'s are Baryon number violating Yukawa couplings.

In the context of the present study, we will work with only $\lambda^{'}_{ijk}$-type of couplings, in particular, with $\lambda^{'}_{112}$ and $\lambda^{'}_{113}$ couplings, purely motivated by the observations of LHCb and CMS. We found that it is the only combination of RPV couplings which can be consistent with both of these observations. 

The coupling constants for the RPV operators are typically small due to the constraints from various observables including proton stability, neutrino mass and mixing, processes with flavor-changing neutral current and CP violation, cosmological baryon asymmetries, etc. (see e.g. Ref.~\cite{Barbier:2004ez} for a comprehensive review). Recently there has been some works for providing an organizing principle that explains why RPV couplings are typically small and hierarchical \cite{Nikolidakis:2007fc, Csaki:2011ge,Csaki:2013jza ,Dreiner:2003hw,Monteux:2013mna}.

The choice of  our RPV couplings $\lambda'_{11k}$ with $k = 2$ and 3, are constrained from various low energy observables such as, (i) charge-current universality, (ii) $e-\mu-\tau$ universality, (iii) atomic parity violation etc. The most stringent bound on individual RPV coupling comes from (i) and (ii) \cite{Kao:2009mz,Kao:2009fg}
\begin{align}
|\lambda'_{11k}|\left(\frac{100\ \text{GeV}}{m_{\tilde{d}_{kR}}}\right)<0.03\, ,
\end{align}
where, $m_{\tilde{d}_{kR}}$ is the mass of the right-handed down-type squark. 

The bounds on the product $|\lambda'_{112}\lambda'_{113}|$ mainly come from charged $B$-meson decay $B^{\pm}_d \rightarrow \pi^{\pm} K^0$, $B_{s}-\bar{B}_{s}$ mixing and $B\rightarrow X_s\gamma$ transition. Assuming the mediator mass to be around 100 GeV these translate to 
\begin{align}
|\lambda'_{112}\lambda'_{113}| \lesssim \begin{cases} 
5.7 \times 10^{-3} & [B^{\pm}_d \rightarrow \pi^{\pm} K^0] \textrm{\cite{Bhattacharyya:1999xh}}, \\
 2.3 \times 10^{-2} & [B_{s}-\bar{B}_{s}] 
 \textrm{\cite{Amhis:2012bh}}  , \\
 3.5 \times 10^{-2} & [B\rightarrow X_s\gamma] \textrm{\cite{deCarlos:1996yh,Dreiner:2013jta}}.
 \end{cases}
\end{align}

In addition to the known bounds on RPV couplings $\lambda'_{11k}$ listed above, we present new bounds obtained by analyzing
the non-observation of ``contact interactions'' from collider searches in the following. 
The collider experiments at the LEP~\cite{Schael:2006wu}, HERA~\cite{Ciesielski:2009zz} and Tevatron~\cite{D0note1,D0note2} have put some bounds \cite{Carpentier:2010ue} on the cut-off scale of the four-fermion operator, $\frac{4\pi}{\Lambda^2_{LR}} (\bar{e}_L\gamma_\mu e_L)(\bar{q}_R\gamma^\mu q_{R})$, with $\{q=s,b\}$: $\Lambda_{LR}\sim 5.2$ TeV for $q=s$ and $\Lambda_{LR}\sim2.8$ TeV for $q=b$. This imposes the following bounds on the RPV couplings:
\begin{align}
|\lambda^{'}_{112} |&< \frac{m_{\tilde{u}_L}}{1.0\ \text{TeV}},\\
|\lambda^{'}_{113} |&< \frac{m_{\tilde{u}_L}}{560\ \text{GeV}}.
\end{align}


Also from a global study of electron-quark contact interactions~\cite{Barger:1997nf,Cheung:1998gg} through ZEUS~\cite{Breitweg:1997ff}, H1~\cite{Adloff:1997fg}, polarized $e^{-}$ on nuclei scattering experiments at SLAC~\cite{Prescott:1979dh}, Mainz~\cite{Heil:1989dz}, and Bates~\cite{Souder:1990ia}, Drell-Yan production at the Tevatron~\cite{Abe:1997gt}, total hadronic cross section $\sigma_{\text{had}}$ at CERN LEP~\cite{Abbaneo:1996pw,Alexander:1996wh,Alexander:1996rg,Ackerstaff:1996sf,Acciarri:1995yr,Acciarri:1997iu,Buskulic:1996ua,Langacker:1997au}, and neutrino-nucleon scattering from CCFR~\cite{McFarland:1997wx}, the highest fit value is found to be $\Lambda_{LR}\sim11.2$ TeV, which translates into
\begin{equation}
|\lambda^{'}_{11q} |< \frac{m_{\tilde{u}_L}}{2.2\ \text{TeV}}.
\end{equation}

\section{Phenomenology}\label{sec:phenos}
\subsection{Lepton non-universality at the LHCb}\label{subsec:Bphysics}

We begin with analyzing the recent result on the measurement of $R_K$ reported by the LHCb collaboration. As we aim for finding a unified framework for explaining two seemingly uncorrelated measurements,
we restrict our analysis for $R_K$ within the context of $eejj$ excess reported by CMS in the context of $W_R$ search only. For this, the RPV $LQD^c$-type $\lambda^{'}_{112}$ and $\lambda^{'}_{113}$ couplings are the most important parameters where the former plays a major role in determining the size of both observables. Therefore, in our $B$-physics analysis we will focus on finding a reasonable parameter space which would allow a sizable $\lambda^{'}_{112}$ coupling compatible with the CMS $eejj$ data. Here we want to emphasize that this particular combination of RPV operators. We found that all the other possible combinations of RPV operator cannot simultaneously explain both CMS and LHCb results. For example, the combination of $\lambda^{'}_{122}$ and $\lambda^{'}_{123}$ can explain the $R_K$, but fails to accommodate the $eejj$ excess due to low slepton production cross-sections. Therefore, the combination of RPV $LQD^c$-type $\lambda^{'}_{112}$ and $\lambda^{'}_{113}$ couplings provide a unique solution to the problems we consider in our paper.

In the SM, $b\rightarrow s$ flavor changing neutral current transition is in general highly suppressed 
due to absence of tree level processes. The leading order contribution comes from electroweak loop processes. Therefore, it provides an important tools to test the flavor sector of the SM, as well as to probe and constrain its possible extensions. In this context, for the exclusive decay ${B}^{+}\rightarrow{K}^{+}ll$ with $l=e,\mu$, one of the very useful observables is the ratio  ($R_K$) of the branching fractions in the individual  lepton flavor mode. 

Theoretically, $R_{K}\approx1$ from the lepton universality in the SM, which ensures that electron and muon couple to the gauge bosons with the same strength. Although the individual branching fractions of ${B}^{+}\rightarrow{K}^{+}ee$ and ${B}^{+}\rightarrow{K}^{+}\mu\mu$  suffer from  theoretical uncertainties of $\mathcal{O}(30\%)$~\cite{Bobeth:2011nj}, $R_K$ remains unaffected as the uncertainties cancel out while taking the ratio~\cite{Hiller:2003js}. Hence, it is a   clean and sensitive observable for probing the extension of the SM, specially for the flavor-non-universality.

The recent measurement of $R_K$ in the low di-lepton invariant mass squared region, $1\ \mathrm{GeV}^2 < q^{2} < 6\ \mathrm{GeV}^{2}$, is found to be \cite{Aaij:2014ora},
\begin{equation}
R_{K}^{\text{LHCb}}=0.745\pm_{0.074}^{0.090}\pm0.036\, .
\end{equation}
This corresponds to a 2.6$\sigma$ deviation from the SM prediction, $R_{K}=1.0003^{+0.00010}_{-0.00007}$ \cite{Hiller:2003js,Bobeth:2007dw}, indicating a possible hint of new physics. As discussed in the introduction, there are two possible explanations: it could be either due to the depletion in $\BR({B}^{+}\rightarrow{K}^{+}\mu\mu)$ or an enhancement in $\BR({B}^{+}\rightarrow{K}^{+}ee)$.
We focus on the latter, in order to explain the CMS $eejj$ excess as well.
 
We begin with considering the following effective weak Hamiltonian for $\bar{s}bll$ transition,
\begin{equation}
\mathcal{H}_{\text{eff}}=-\frac{4G_{F}}{\sqrt{2}}V_{tb}V_{ts}^{*}\frac{\alpha_{e}}{4\pi}\sum_{i}C_{i}(\mu)\mathcal{O}_{i}(\mu),
\end{equation}
where $\alpha_{e}$, $V_{ij}$, $G_{F}$ and $\mu$ are the fine structure constant, the CKM matrix elements, the Fermi constant, and the renormalization scale respectively.
The relevant dimension six $\bar{s}bll$ operators in our case are vector and axial-vector operators
\begin{equation}
\mathcal{O}_{9}=(\bar{s}\gamma_{\mu}P_{L}b)(\bar{l}\gamma^{\mu}l),\;\;\;\;
\mathcal{O}_{10}=(\bar{s}\gamma_{\mu}P_{L}b)(\bar{l}\gamma^{\mu}\gamma_{5}l).
\end{equation}
The corresponding chirality-flipped operators $\mathcal{O}'$ are obtained by changing $P_{L}\leftrightarrow P_{R}$.
It is convenient to divide the Wilson coefficients as
\begin{equation}
C^{(\prime)}(\mu)=C^{\text{SM}(\prime)}(\mu) + C^{\text{NP}(\prime)}(\mu),
\end{equation}
where, $C^{\text{SM}(\prime)}(\mu)$ is the SM contribution and $C^{\text{NP}(\prime)}(\mu)$ is the NP contribution. We have $C_{9}^{\text{SM}}(m_b)=-C_{10}^{\text{SM}}(m_b)=4.2$ for all leptons while rest of the semileptonic Wilson coefficients are negligible \cite{Bobeth:2012vn}. 

\begin{figure}[t]
\centering
\includegraphics[width=0.48\textwidth]{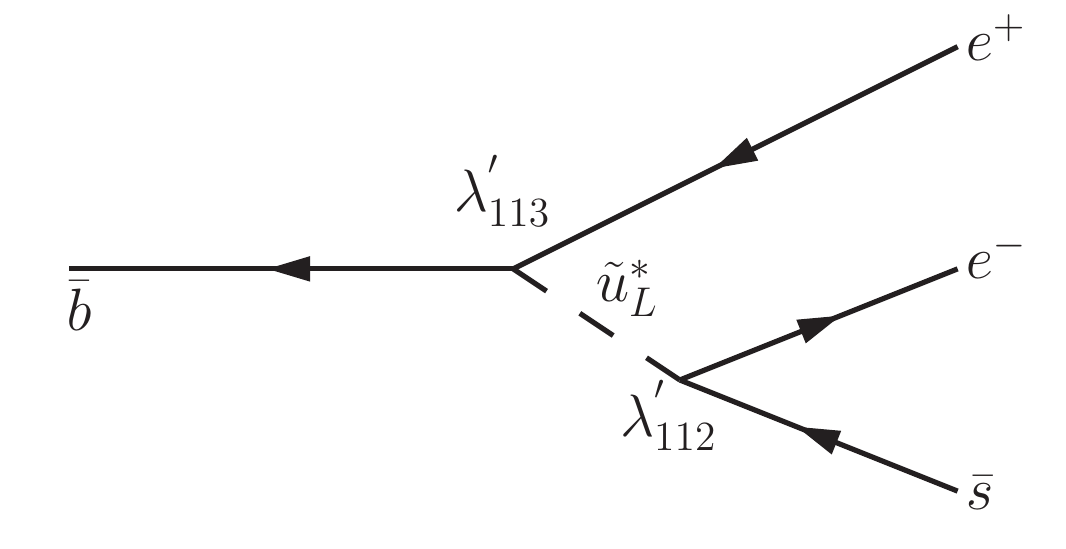}
\caption{Feynman diagram of $\bar{b}\rightarrow\bar{s}e^{+}e^{-}$ decay via $\lambda^{'}_{113}$ and $\lambda^{'}_{112}$  coupling.}
\label{fig:tree}
\end{figure}

As described in section~\ref{sec:intro}, in this work we consider the following \R-parity violating term in Eq.~(\ref{eq:WRPV}) as a source of NP,
\begin{equation}
\mathcal{L}\ni \lambda'_{ijk}L_{i}Q_{j}D^{c}_{k}\, .
\end{equation}
In the context of $\bar{b} \to \bar{s} e e$ transition (Fig. \ref{fig:tree}), we analyze the coupling between up-type squark $\tilde{u}_L$, down-type quarks $s$ and $b$, and electron,
\begin{equation}
\mathcal{L}\ni \lambda'_{112}\tilde{u}_{L}(\bar{s}P_{L}e) +\lambda_{113}'\tilde{u}_{L}(\bar{b}P_{L}e) + \mathrm{h.c.}\, .
\end{equation}
Integrating out $\tilde{u}_L$, we obtain the following effective Hamiltonian
\begin{align}
\mathcal{H}_{\text{eff}} =& -\frac{\lambda'_{112}\lambda_{113}'^{*}}{m_{\tilde{u}_L}^{2}}(\bar{s}P_{L}e)(\bar{e}P_{R}b)\nonumber\\
=&\frac{\lambda'_{112}\lambda_{113}'^{*}}{2m_{\tilde{u}_L}^{2}}(\bar{s}\gamma^{\mu}P_{R}b)(\bar{e}\gamma_{\mu}P_{L}e)\nonumber\\
=&-\frac{4G_{F}}{\sqrt{2}}V_{tb}V_{ts}^{*}\frac{\alpha_{e}}{4\pi}(C_{9}'^{e}\mathcal{O}_{9}'^{e}+C_{10}'^{e}\mathcal{O}_{10}'^{e}),
\end{align}
where, $m_{\tilde{u}_L}$ is the mass of $\tilde{u}_L$, and the Wilson coefficients in terms of the RPV operators are given by
\begin{equation}
\label{eq12}
C_{10}'^{e}=-C_{9}'^{e}=\frac{\lambda_{112}'\lambda_{113}'^{*}}{V_{tb}V_{ts}^{*}}\frac{\pi}{\alpha_{e}}\frac{\sqrt{2}}{4m_{\tilde{u}_L}^{2}G_{F}}.
\end{equation}
For simplicity, we will suppress the explicit $\mu$ dependencies of the Wilson coefficients from here onward. Since we only have the vector and axial-vector operators, it is straightforward to obtain bounds on the relevant parameters from the experimental data.

Here, following the leptoquark model in \cite{Hiller:2014yaa}, we focus on fitting $R_K$ exclusively, allowing other observables (e.g. $B\rightarrow K^* l^+l^-$ \cite{Bobeth:2012vn,Aaij:2012vr}, $B\rightarrow X_s l^+l^-$ \cite{Huber:2005ig,Lees:2013nxa}, $B\rightarrow e^+e^-$ \cite{Bobeth:2013uxa,CMSandLHCbCollaborations:2013pla}, and $B\rightarrow \mu^+\mu^-$ \cite{Bobeth:2013uxa,Beringer:1900zz}), affected by the same operators, to be consistent within $1 \sigma$ region \cite{Ghosh:2014awa}. 
It is worth noting that several inclusive analyses on these operators, specially in the context of angular observables in $B\rightarrow K^* \mu^+\mu^-$, have been performed \cite{Alonso:2014csa,Descotes-Genon:2013wba,Altmannshofer:2013foa,Beaujean:2013soa,Horgan:2013pva}, as well as a recent global fit with inclusion of $R_K$ \cite{Ghosh:2014awa}. While current global fit prefers NP to be appearing in left-handed 
current with muon, rather than in left-handed current operator with electron we are considering, the tension is only within
$1 \sigma$ range. Therefore, in light of explaining $R_K$ from NP contribution, it is not unreasonable to 
consider operators involving left-handed electron current as was done in the leptoquark model \cite{Hiller:2014yaa}\footnote{ 
For detailed discussion of other $B$ physics observables, see the discussion in~\cite{Hiller:2014yaa}, specially for section III-A 
for (Axial)-vectors operator and section IV-A for a model with a $RL$ operator for electrons. Note that, as mentioned in the 
introduction, this model is just the same as ours except that it's on third generation, and hence cannot explain the CMS excess.}. In addition, when one also addresses the CMS $eejj$ excess at the same time, this choice becomes inevitable.


 
Following \cite{Das:2014sra,Hiller:2014yaa}, the bound on the Wilson coefficients coming from $R_K$ at $1\sigma$ level is given by
\begin{align}\label{const1-1}
0.7&\lesssim\text{Re}[X^{e}-X^{\mu}]\lesssim 1.5\, , 
\end{align}
where, $X^{e}=2C_{9}'^{e}$ and $X^{\mu}=0$ in our case. 

The other important bound comes from the $\bar{B}_{s}$ decay.
In the absence of the scalar and pseudoscalar operator, the model independent constraint is given by \cite{Hiller:2014yaa},
\begin{equation}
\frac{\BR(\bar{B}_{s}\rightarrow ee)^{\rm NP}}{\BR(\bar{B}_{s}\rightarrow ee)^{\text{SM}}}=|1 + 0.24\, C_{10}'^{e}|^{2}.
\end{equation}
The corresponding experimental data\cite{Beringer:1900zz} and SM value\cite{Bobeth:2013uxa} of $\BR(\bar{B}_{s}\rightarrow ee)$, and their ratio are given by
\begin{align}
\BR(\bar{B}_{s}\rightarrow ee)^{\text{exp}} <&\ 2.8\times10^{-7},\\
\BR(\bar{B}_{s}\rightarrow ee)^{\text{SM}} =&\ (8.54\pm0.55)\times10^{-14},\\
\frac{\BR(\bar{B}_{s}\rightarrow ee)^{\text{exp}}}{\BR(\bar{B}_{s}\rightarrow ee)^{\text{SM}}} <&\ 3.3\times 10^{6}.
\end{align}
From these, we obtain a bound on the Wilson coefficient $C_{10}'^{e}$, 
\begin{equation}
\left|1+0.24\cdot C_{10}'^{e}\right|^{2}<3.3\times10^{6}.\label{const1-2}
\end{equation}
Plugging in the values of input parameters \cite{Beringer:1900zz}, $|V_{tb}|=0.999146^{+0.000021}_{-0.000046}$,
$|V_{ts}|=0.0404^{+0.0011}_{-0.0005}$,
$G_{F}=1.166 378 7\times10^{-5}\text{GeV}^{-2}$, and $\alpha_{e}(m_{b})=1/133$, in Eq.~(\ref{eq12}), the Wilson coefficient $C_{9, 10}'^{e}$ become
\begin{equation}
C_{10}'^{e}=-C_{9}'^{e}\simeq-\frac{\lambda_{112}'\lambda_{113}'^{*}}{2m_{\tilde{u}_L}^{2}}(21\ \text{TeV})^{2}.
\label{C10C9}
\end{equation}
Substituting this into Eq.~(\ref{const1-1}) and Eq.~(\ref{const1-2}), we obtain the following constraints on our model parameters, $m_{\tilde{u}_L}, \lambda'_{112}$ and $\lambda'_{113}$,
\begin{align}
\frac{0.7}{(21\ \text{TeV})^{2}}\lesssim\text{Re}\left[\frac{\lambda_{112}'\lambda_{113}'^{*}}{m_{\tilde{u}_L}^{2}}\right]\lesssim \frac{1.5}{(21\ \text{TeV})^{2}},\\
\left|1-0.12\frac{\lambda_{112}'\lambda_{113}'^{*}}{m_{\tilde{u}_L}^{2}}(21\ \text{TeV})^{2}\right|^{2}<3.3\times10^{6}.
\label{equ:bound}
\end{align}
The above two equations sets the hierarchy between the two couplings $\lambda_{112}'$ and $\lambda_{113}'$ for a fixed value of $m_{\tilde{u}_L}$.

\subsection{Lepton non-universality at the CMS}\label{subsec:Collider}

\begin{figure}[t]
\centering
\includegraphics[width=0.48\textwidth]{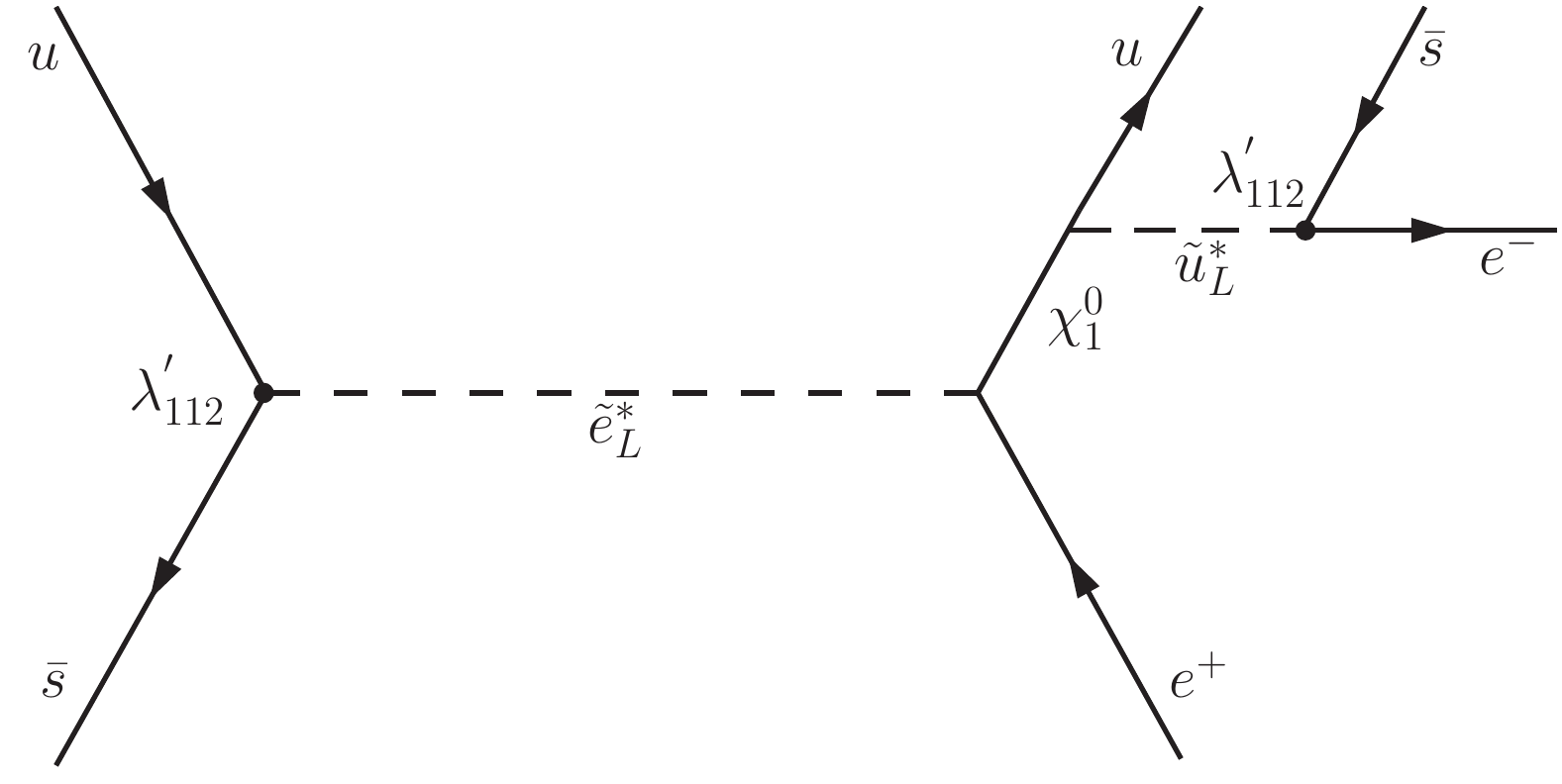}
\includegraphics[width=0.48\textwidth]{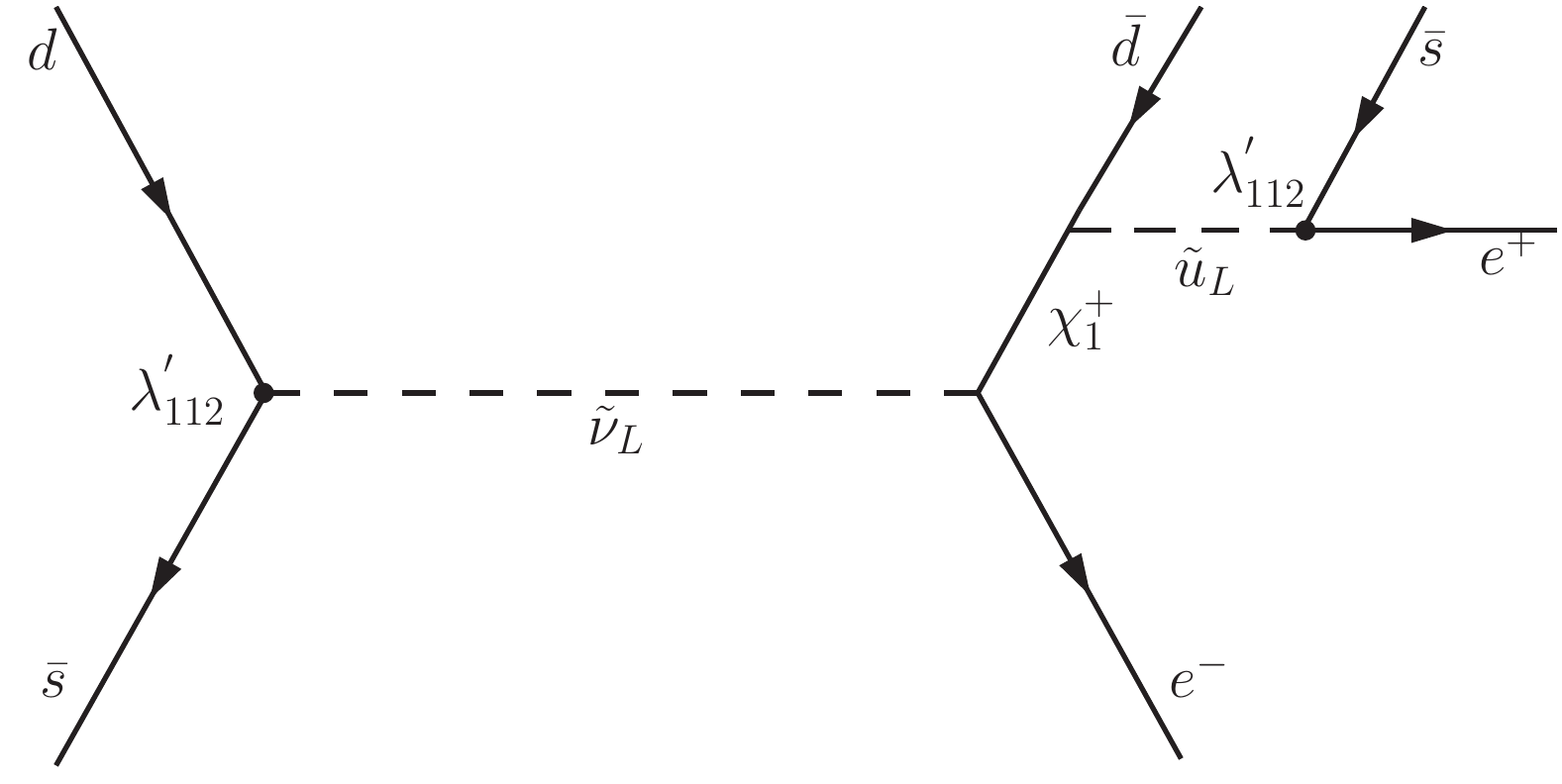}
\caption{Feynman diagram of (a) resonant selectron production (left) and (b) resonant sneutrino production via $\lambda^{'}_{112}$ coupling.}
\label{fig:feyndia}
\end{figure}

We have considered resonant slepton production via $\lambda'_{112}$ coupling in $pp$ collision
at the LHC at 8 TeV center of mass energy with 19.7 \fb integrated luminosity. 
The resonant slepton production at collider experiments has been studied extensively in \cite{Dimopoulos:1988fr,Hewett:1998fu,Dreiner:1998gz,Dreiner:2000qf,Dreiner:2000vf,Allanach:2003wz,Dreiner:2012np,Allanach:2009iv}.
The slepton thus produced can decay via both $\mathcal{R}$-parity conserving and violating couplings. 
The branching ratio  depends on the mass of the lightest SUSY particle (LSP) and the $\lambda'_{112}$ coupling. Both the
selectron and the sneutrino have a substantial decay branching fraction of decaying into $e \chi_1^0$ 
and $e \chi_1^+$ respectively. The lightest neutralino and lighter chargino can further decay via the RPV coupling 
resulting in a $eejj$ final state as studied in the context of the $W_R$ search at CMS \cite{Khachatryan:2014dka},
\begin{align}
p p \rightarrow& \tilde{e}_L \rightarrow e \chi^0_1 \rightarrow e e j j\, , \\
p p \rightarrow& \tilde{\nu}_L \rightarrow e \chi^+_1 \rightarrow e e j j\, .
\end{align}
In Fig.~\ref{fig:feyndia} we show the Feynman diagrams leading to the above final state through the resonant production of selectron (left) and sneutrino (right). 
We have considered three different benchmark scenarios to take into account the model dependency
in the branching ratio calculation:
\begin{itemize}
\item {\bf Bino-like scenario:} $M_1 \ll M_2 $, the LSP is dominated by the bino-component, with heavy wino mass 
($> 2$ TeV). In this scenario the branching ratio of the slepton decay via $\mathcal{R}$-parity conserving
coupling can be subdominant compared to the di-jet mode.

 \item {\bf Mixed scenario:} $M_1:M_2 = 2:3$, {i.e.}, the LSP is mostly bino-like with a small wino-admixture. In this case the slepton has a substantial branching ratio of decaying into second lightest neutralino ($\chi_2^0$) and lighter chargino ($\chi_1^{+}$). Both the $\chi_2^0$ and $\chi_1^{+}$ in this case decay via $\mathcal{R}$-parity conserving coupling, hence reducing the effective $\tilde{\ell} (\tilde{\nu})\rightarrow eejj$ branching ratio.

\item {\bf Mixed inverted scenario:} $M_1:M_2 = 3:2$, {i.e.}, the LSP is mostly wino-like with a small bino-admixture. 
In this case the slepton has a substantial branching ratio of decaying to the lightest neutralino and lighter chargino. 
In this scenario, however, both the lighter chargino and the lightest neutralino decay via RPV coupling. Hence, the lepton and 
jet multiplicity get enhanced in the final state compared to the above two cases.

\item {\bf Wino-like scenario:} $M_2 \ll M_1$, {i.e.}, the LSP is purely wino-like. This scenario is similar to above
(mixed-inverted) one with a slight enhancement in the effective $\tilde{\ell} (\tilde{\nu})\rightarrow eejj$ branching
ratio.
\end{itemize}
\begin{figure}[t]
  \includegraphics[width=0.5\textwidth]{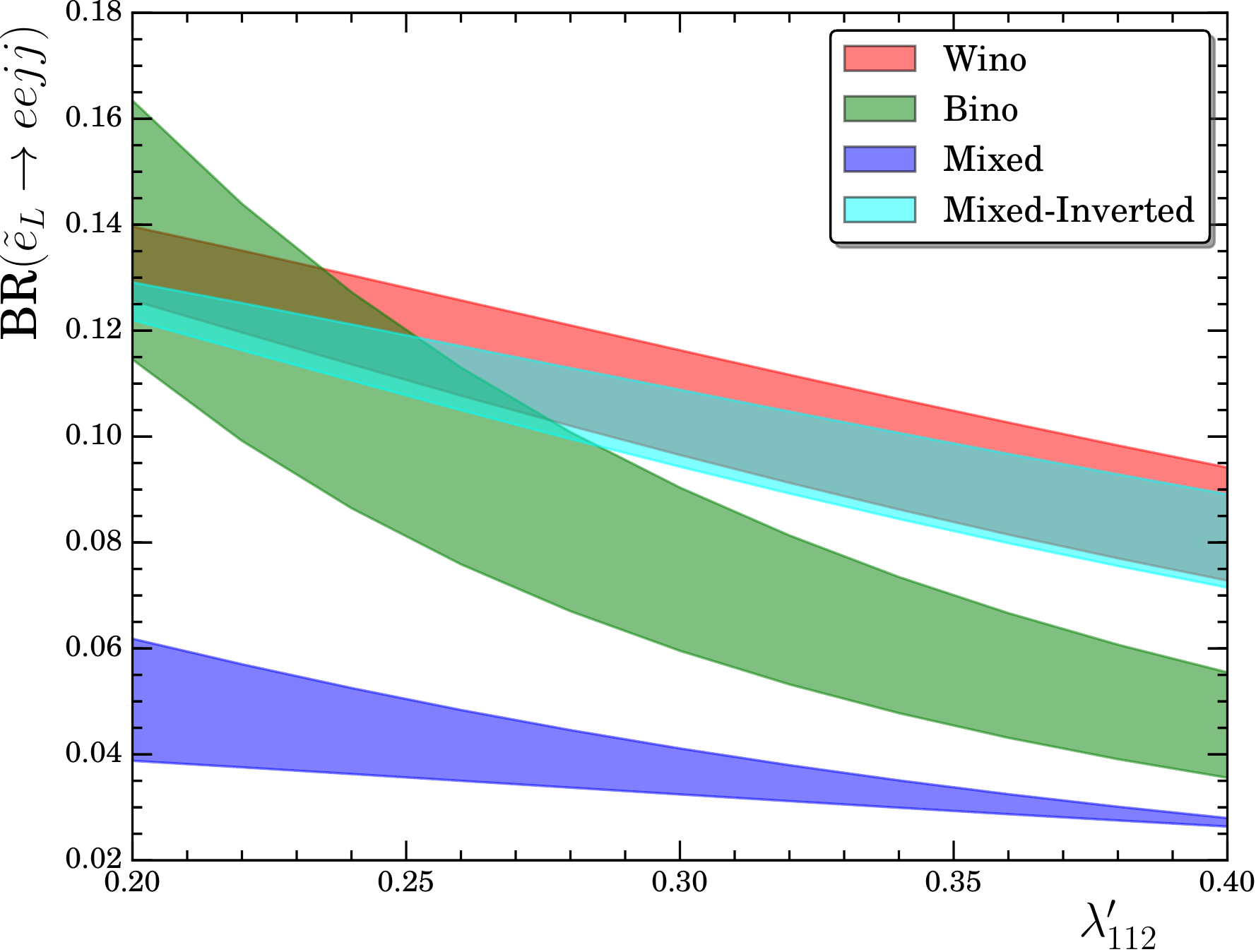}
  \caption{The effective BR$(\tilde{e}_L \to e e jj)$ vs. $\lambda'_{112}$ plot for bino, mixed, mixed-inverted and wino-like scenarios. Each color band corresponds to the neutralino mass varied within the range [300, 1000] GeV.}
  \label{fig:br3}
\end{figure}

The model spectrum and decay branching ratios have been calculated using SARAH-4.3.1 \cite{Staub:2009bi,Staub:2010jh} and SPheno-3.3.2 \cite{Porod:2003um,Porod:2011nf}. In Fig. \ref{fig:br3} we present the effective $\tilde{e}_L \to e e jj$ branching ratio vs. $\lambda^{\prime}_{112}$. For the rest of the work we will assume the lighter slepton masses of the first generation $m_{\tilde{e}_L,\tilde{\nu}_L} = 2.1$ TeV and dominantly left-chiral. Squark masses of first generation are $\sim 1.5$ TeV and rest of the sfermions are set at higher values than these. We vary the lightest neutralino mass in the range [300, 1000] GeV. The  bound coming from the narrow di-jet resonance search by CMS \cite{Chatrchyan:2013qha} on the $\sigma\times$Br$(\tilde{l}\to j j)\times \mathcal{A}$, where, $\mathcal{A}$ is the efficiency of cut, is 45 \fb for a resonant mass around 2.1 TeV. The choice of our coupling ranges ($0.2 \leq\lambda^{'}_{112} \leq 0.4$) are consistent with the above bound.


We have simulated the resonant slepton production in $pp$ collision at the LHC using MadGraph5 \cite{Alwall:2011uj} and the subsequent decays, initial and final state radiation, parton showering and hadronization effects have been done using PYTHIA (v6.4) \cite{Sjostrand:2006za}. We have worked with CTEQ6L \cite{Kretzer:2003it} parton distribution function. The factorization and the renormalization scales are set at the slepton mass $\mu_F=\mu_R={m}_{\tilde{e}_L}$. To take into account the next-to-leading order QCD correction we multiply the tree-level cross-section by the $K$-factor 1.34 \cite{Dreiner:2012np}. We have also used various resolution functions parametrized as in \cite{Chatrchyan:2011ds} for the final state objects to model the finite detector resolution effects. We have assumed the object definition described in \cite{Khachatryan:2014dka} for the final state particles along with the following cuts,
\begin{itemize}
\item Invariant mass of the electron pair, $M_{ee}> 200$ GeV.
\item Invariant mass of the two electrons and dijet system, $M_{eejj}> 600$ GeV.
\end{itemize}

\section{Results and Discussion}\label{sec:conclusions}

\begin{table}[t]
   \begin{center}
  \begin{tabular}{l c c c } 
    \hline \hline
     Cut  & {Signal} & Background & Data \\
    \hline
    $2e+\geq 2j$  &    11.72    &  34154  &   34506   \\
    $M_{ee}> 200$ GeV  &  11.71     &   1747  &   1717  \\
    $M_{eejj}> 600$ GeV  &   11.71       &   $783\pm51$  &  817   \\
    1.8 TeV $< M_{eejj} <$ 2.2 TeV  &  10.01     &  $4.0\pm1.0$   &   14   \\
    \hline \hline
  \end{tabular}
\caption{Number of events from signal, backgrounds and  reconstructed data  assuming $\lambda'_{112}=0.22$, $m_{\tilde{e}_L}=2.1$ TeV and $m_{{\chi}^0_1}=400$ 
GeV in the wino-like benchmark scenario at 19.7 \fb integrated luminosity and 8 TeV center of mass energy.  The data, SM backgrounds and selection cuts are taken from ref. \cite{Khachatryan:2014dka}.}
  \label{tab:events}
   \end{center}
\end{table} 
\begin{figure}[t]
  \includegraphics[width=0.5\textwidth]{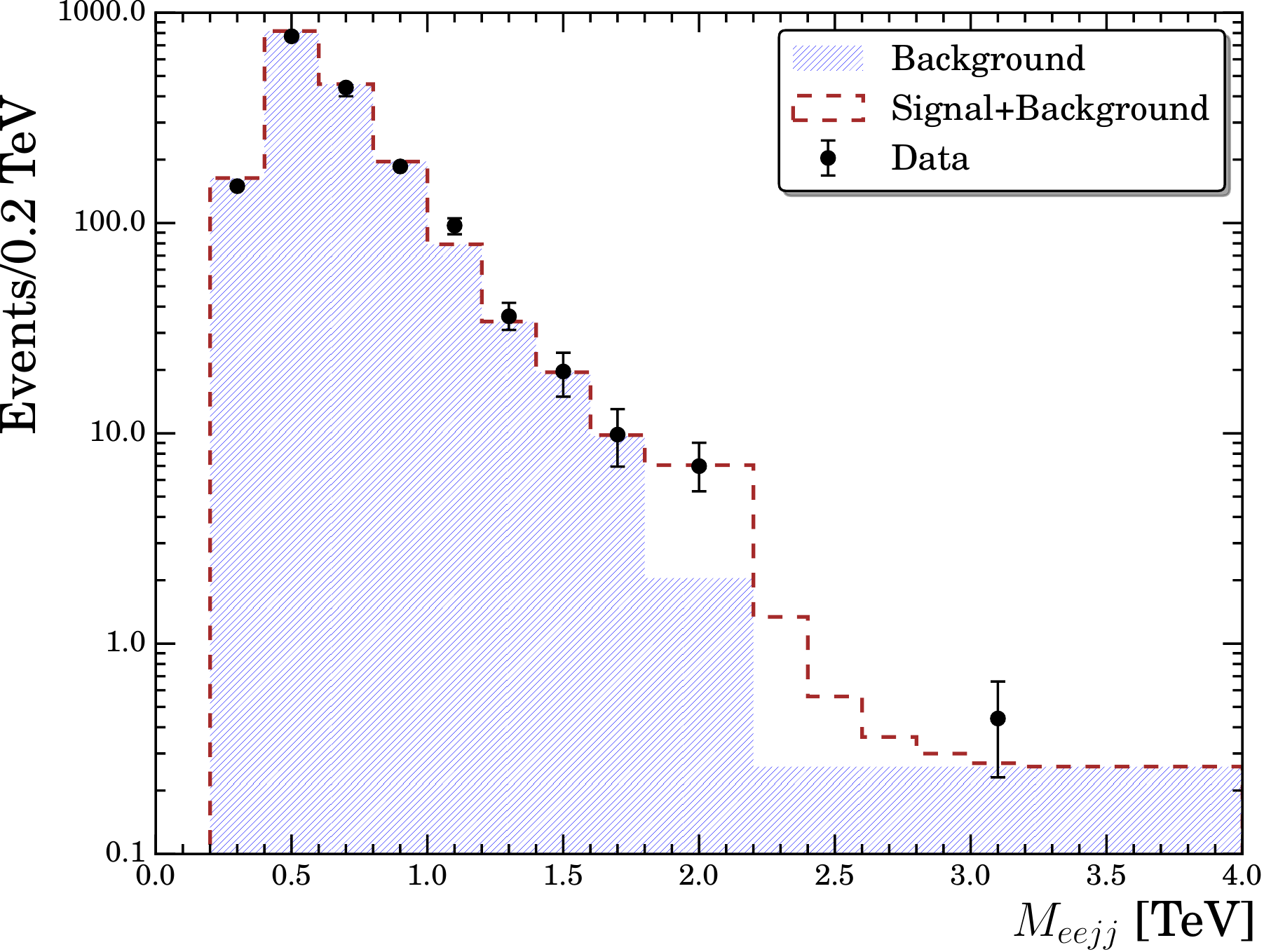}
\caption{{A comparison of the data, signal and background of the $M_{eejj}$ distributions after imposing cuts used in analysis of the $W_R$ search. The signal point corresponds to $\lambda'_{112}=0.22$,  $m_{\tilde{e}_L}=2.1$ TeV and $m_{{\chi}^0_1}=400$ GeV in the wino-like benchmark scenario. The data and SM backgrounds are taken from \cite{Khachatryan:2014dka}. }}
  \label{fig:Meejj}
\end{figure}

\begin{figure}[t]
  \includegraphics[width=0.48\textwidth]{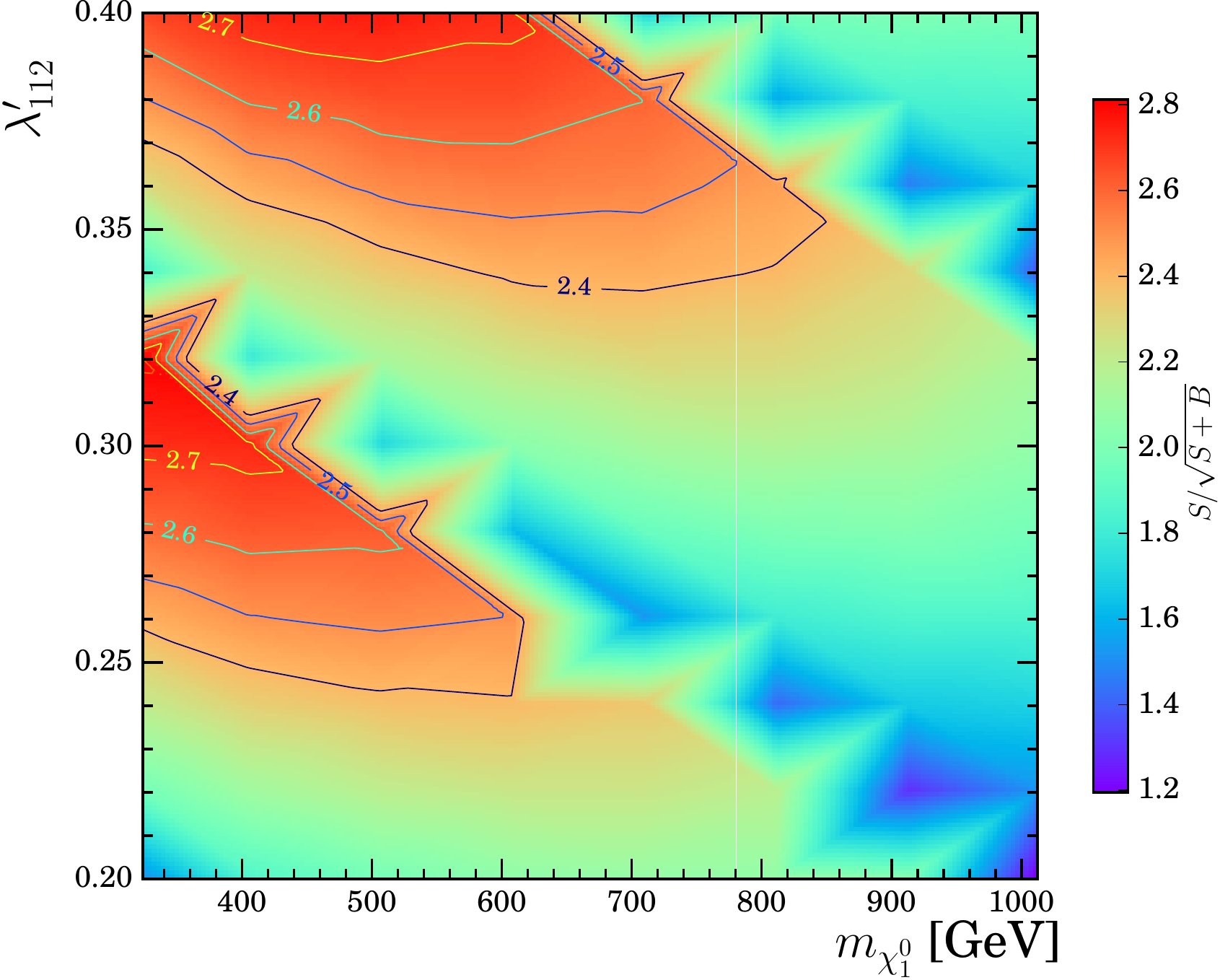}
  \includegraphics[width=0.48\textwidth]{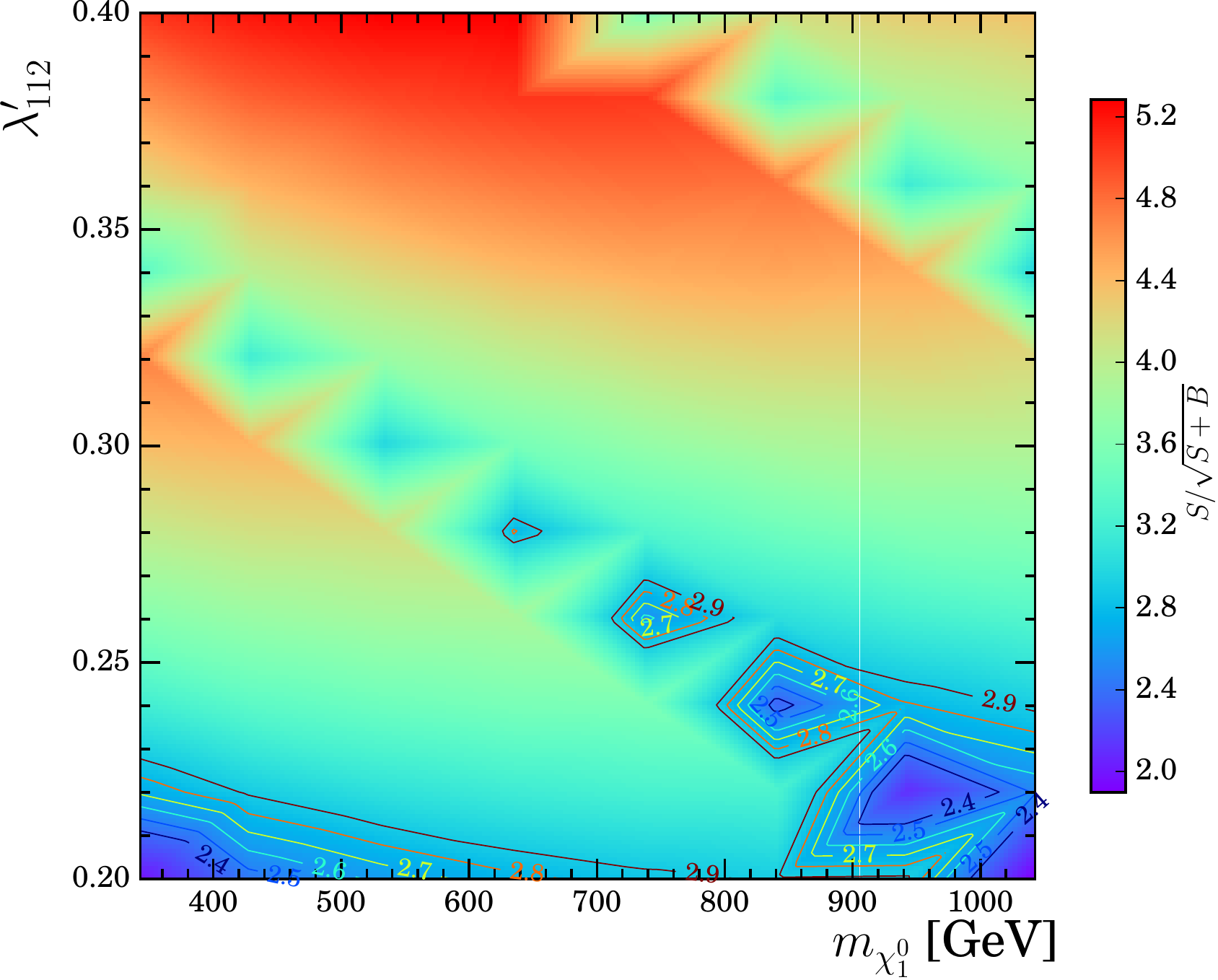}
  \caption{The region of $\lambda'_{112}$-$m_{{\chi}^0_1}$ parameter space compatible with the CMS excess ($W_R$ search) for the (a) bino (left) and (b) wino-like (right) scenario. The color gradient signify the $S/\sqrt{S+B}$ estimate of the signal where, $S$ is the signal event and $B$ is the background event within $1.8$ TeV $\leq M_{eejj} \leq 2.2$ TeV. It is not possible to explain the CMS excess assuming the mixed scenario due to a very small effective branching fraction BR$(\tilde{e}_L \to e e jj)$. The mixed-inverted scenario also gives similar excess as in wino-like scenario. }
\label{fig:param-bino-wino}
\end{figure}

In this article, we have addressed the recent CMS and LHCb results from a unified framework. The results of our analysis are shown in 
Table \ref{tab:events} and Fig. \ref{fig:Meejj} which show the comparison of signal, background and 
the corresponding data for a typical benchmark point of wino-like scenario. A more detailed  
analysis depicting the range of $\lambda^{\prime}_{112}$ coupling which can be compatible with 
the CMS result is presented in Fig. \ref{fig:param-bino-wino}. 
From Fig. \ref{fig:param-bino-wino}a, one can see that there are two distinct regions corresponding to low values of $m_{\chi_1^0}$ where one does have $2.8\sigma$ significance in the bino-like scenario. The color gradient in Fig. \ref{fig:param-bino-wino} signifies the $S/\sqrt{S+B}$ estimate\footnote{The significance defined here is different from that used by the CMS collaboration. However, we work with this simple definition to find the potential region of parameter space that can explain the CMS excess.} of the signal where, $S$ is the signal event and $B$ is the background event within $1.8$ TeV $\leq M_{eejj} \leq 2.2$ TeV. This is due to the fact that, the
cross-section grows with $|\lambda^{\prime}_{112}|^2$, whereas the $\mathcal{R}$-parity conserving 
decay branching ratios of the slepton falls with the increase in $\lambda^{\prime}_{112}$ (see Fig. \ref{fig:br3}). 
Thus two regions have been obtained which give equal event rates in the $eejj$ analysis. For the bino-like scenario and the mixed scenario the contribution to $eejj$ 
final state mainly comes from resonant selectron production (Fig. \ref{fig:feyndia}a). The contribution from the resonant sneutrino production is negligible due to the fact that the RPV decay of the chargino (see Fig. \ref{fig:feyndia}b) has a very small branching ratio.

We can see from Fig. \ref{fig:param-bino-wino}b the allowed value of $\lambda^{'}_{112}$ is smaller compared to the bino-like case owing to the fact that additional contribution coming from resonant sneutrino production and also enhanced effective branching ratio of BR$(\tilde{e}_L \to e e jj)$ compared to the bino-like scenario. Note that both the low and high values of the neutralino mass for a $\lambda^{'}_{112} \sim 0.21$ give same excess compatible with the CMS result. This can be explained by the fact that the reduced branching ratio for $m_{{\chi}^0_1} \sim 1$ TeV is compensated with a higher cut-efficiency.

We emphasize here that the CMS excess has reported
the data summed over bins having total width of 400 GeV. The distribution of the data within this range is
yet unknown. A fine binning of data is required at high luminosity run. In case the data is distributed over this wide range, a resonance explanation of a given mass may not be a good option. However, the wino and mixed-inverted scenario can be better suited in such a case. This requires a splitting $\mathcal{O}(10^2 ~{\rm GeV})$ between the selectron and the sneutrino which can be achieved by introducing large RPV soft-terms of the same type. 

\begin{figure}[t]
\centering 
\subfigure{
\includegraphics[width=0.48\textwidth]{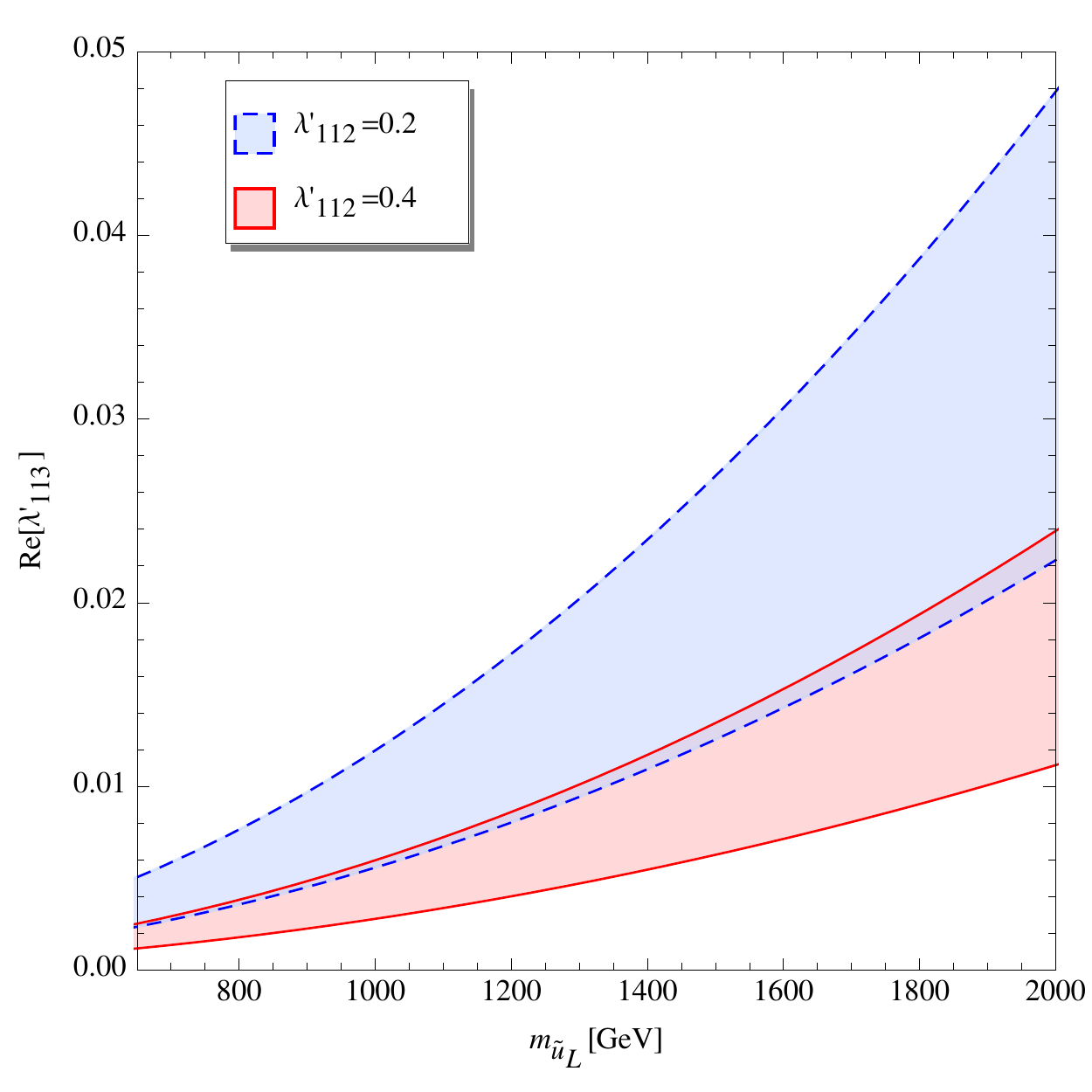}}
\caption{The allowed parameter space in the $m_{\tilde{u}_L} - \text{Re}[\lambda'_{113}]$ plane consistent with the measurement of $R_K$ at LHCb and other $B$-physics observations discussed in section \ref{subsec:Bphysics}. The blue (red) colored region corresponds to $\lambda'_{112}=0.2$ ($0.4$).}
\label{fig:suppara}
\end{figure}

The CMS $eejj$ excess constrains the $\lambda^{'}_{112}$ coupling independent of the LHCb result. We use the results discussed above to constrain the parameter space of $B$-physics analysis, namely, the $\lambda^{'}_{113}$ coupling and relevant mass parameter $(m_{\tilde{u}_L})$. Fig. \ref{fig:suppara} shows plot in the $\lambda^{'}_{113}$-$m_{\tilde{u}_L}$ plane consistent with the experimental
data coming from the measurements listed in section \ref{subsec:Bphysics} assuming two fixed values of 
$\lambda^{'}_{112}=0.2$ and 0.4. 
The present LHC bound on $m_{\tilde{u}_L}$ ($> 1$ TeV) in presence of $\lambda^{\prime}$ couplings comes from the di-leptoquark search
analysis \cite{CMS:2014qpa}. This leaves us with the choice of $\lambda^{'}_{113}$ as low as 0.006 (0.0125) for $\lambda^{'}_{112}=0.4 ~(0.2)$.

In summary, our important observation is that the RPV SUSY operator which can explain the lepton non-universality hinted by the measurement 
of $R_K$ from LHCb, can easily accommodate the lepton non-universality observed by CMS in the context of $W_R$ search. 
In this analysis we do not address the CMS $eejj$ and $e\nu jj$ excesses in the context of di-leptoquark search. A dedicated analysis is presented in
\cite{Allanach:2014nna} to show that the CMS di-leptoquark result can be accommodated within this framework as well. We note that, future measurements in all these sectors can tell us with certainty whether the current deviations are robust or not. As an outlook, we also suggest few collider signatures such as, lepton charge asymmetry measurement in the $e\nu jj$-channel and ratio of same-sign di-lepton events to opposite-sign di-lepton events in the $eejj$ channel to further discriminate our scenario at the LHC. Our result in the context of $R_K$ will be confronted with all the other $B$-physics observables, which might further constrain the allowed range of the parameter space of the model considered here. The effective operators considered here may also give rise to rare $B$-decays like $b \to s \nu \bar{\nu}$ \cite{Buras:2014fpa}, which could be a promising channel for future $B$-physics experiments.


\bigskip
\section*{Acknowledgements} 
We would like to thank Luca Silvestrini for useful inputs. SL would like to express a special thanks to the Mainz Institute for Theoretical Physics (MITP) for its hospitality and support. SL is also grateful to the Dipartimento di Fisica, Universit\`a di Roma La Sapienza for its hospitality during the completion of this work. SB would like to thank Subhadeep Mondal regarding computational help. DC is supported by the European Research Council under the European Union's Seventh Framework Programme (FP/2007-2013)/ERC Grant Agreement n. 279972. SL is supported by the National Research Foundation of Korea(NRF) grant funded by the Korea government(MEST) (No. 2012R1A2A2A01045722).
\bigskip


\bibliographystyle{utphys}
\bibliography{reference}


\end{document}